\documentclass[a4paper]{article}

\usepackage{INTERSPEECH2020}
\usepackage{amsmath,graphicx}
\usepackage{tikz,tikz-qtree}
\usepackage{fix-cm}
\usepackage{booktabs} 
\usepackage{makecell} 
\usepackage{mathtools} 
\PassOptionsToPackage{hyphens}{url}\usepackage{hyperref}
\usepackage{tipa}
\usepackage{lipsum}
\usepackage{url}

\newcommand\blfootnote[1]{%
  \begingroup
  \renewcommand\thefootnote{}\footnote{#1}%
  \addtocounter{footnote}{-1}%
  \endgroup
}

\title{Coswara - A Database of Breathing, Cough, and Voice Sounds for COVID-19 Diagnosis}

\name{Neeraj Sharma, Prashant Krishnan, Rohit Kumar, Shreyas Ramoji, Srikanth Raj Chetupalli, \\Nirmala R., Prasanta Kumar Ghosh, and Sriram Ganapathy}
\address{Indian Institute of Science, Bengaluru, India, 560012}
\email{sriramg@iisc.ac.in}

\begin{document}
\maketitle
\begin{abstract}
The COVID-19 pandemic presents global challenges transcending boundaries of country, race, religion, and economy. The current gold standard method for COVID-19 detection is the reverse transcription polymerase chain reaction (RT-PCR) testing. However, this method is expensive, time-consuming, and violates social distancing. Also, as the pandemic is expected to stay for a while, there is a need for an alternate diagnosis tool which overcomes these limitations, and is deployable at a large scale. The prominent symptoms of COVID-19 include cough and breathing difficulties. We foresee that respiratory sounds, when analyzed using machine learning techniques, can provide useful insights, enabling the design of a diagnostic tool. Towards this, the paper presents an early effort in creating (and analyzing) a database, called Coswara, of respiratory sounds, namely, cough, breath, and voice. The sound samples are collected via worldwide crowdsourcing using a website application. The curated dataset is released as open access. As the pandemic is evolving, the data collection and analysis is a work in progress. We believe that insights from analysis of Coswara can be effective in enabling sound based technology solutions for point-of-care diagnosis of respiratory infection, and in the near future this can help to diagnose COVID-19.
\end{abstract}

\ninept

\section{Introduction}
\begin{figure}[t!]
    \centering
    \includegraphics[width=\linewidth]{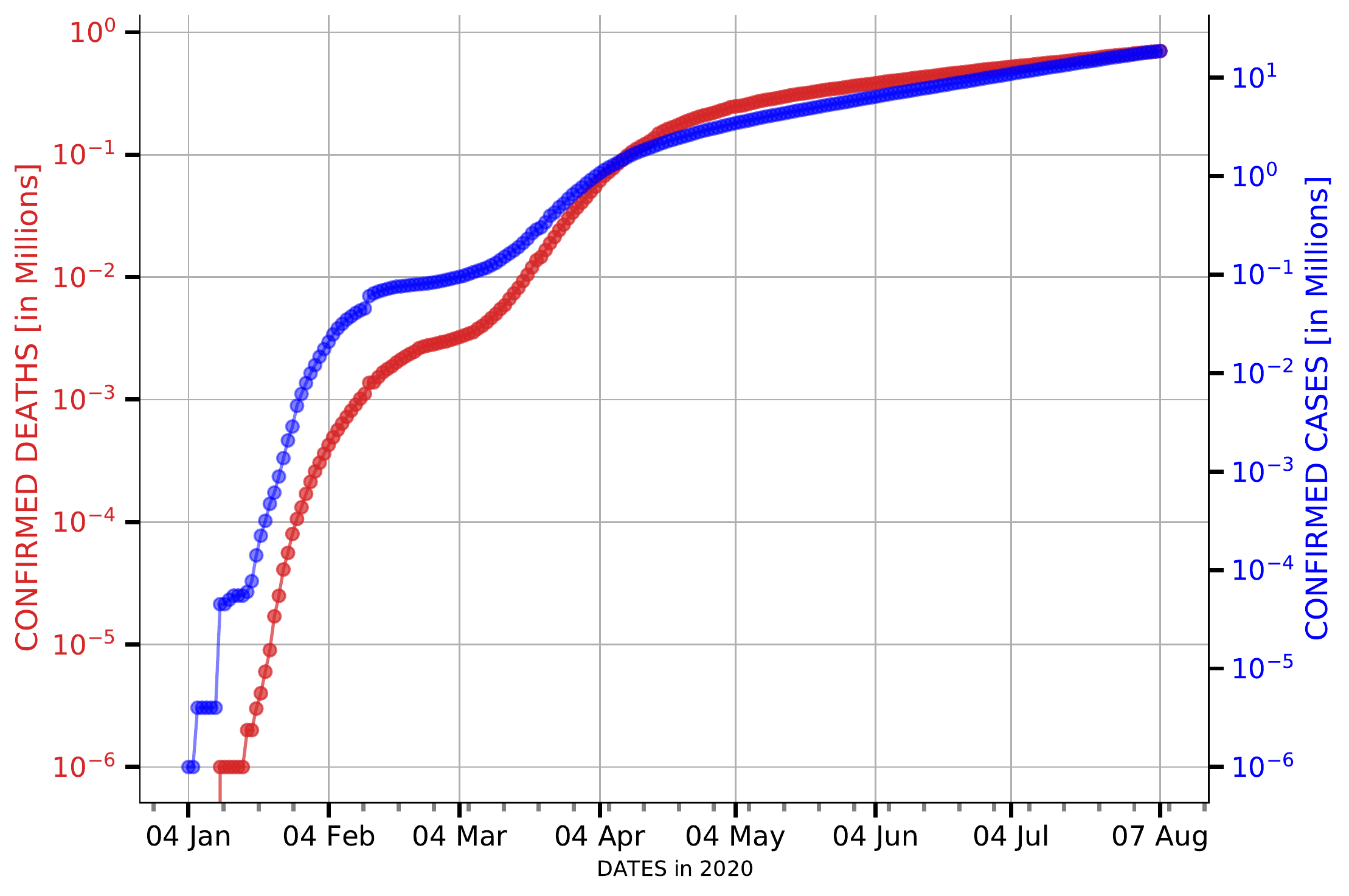}
    \caption{Evolution of COVID-19 cases and deaths. The data is obtained from the WHO Coronavirus Disease (COVID-19) Dashboard on 07 August 2020.}
    \label{fig:covid19_who}
    \vspace{-0.4cm}
\end{figure}

The COVID-19 is a respiratory infection caused by severe acute respiratory syndrome coronavirus $2$ (SARS-CoV-2)~\cite{world2020naming}. The disease, officially declared a pandemic, has infected  millions of humans across the globe, and has a fatality rate between $1-10$\% in most countries. Fig.~\ref{fig:covid19_who} shows the cumulative number of cases (and casualties) as of August $7$, $2020$, and there is still no sign of flattening.
This trajectory of growth started on 4 Jan 2020, and has forced many countries to take serious containment measures such as nation-wide lockdowns and scaling up of the isolation facilities in hospitals. The lockdown is useful as it gives time for large scale testing of individuals. The gold standard for COVID-19 diagnosis is the reverse transcription polymerase chain reaction (RT-PCR) of infected secretions (from nasal or throat cavity).
The results of a RT-PCR test are available in $2-48$ hours. The limitations of the testing include: $(i)$ violation of social distancing which increases the chance of infection spread, $(ii)$ expenses involved in the chemical reagents and devices, $(iii)$ testing time in hours and needs expertise, and $(iv)$ difficulty in large scale deployment. Foreseeing a rise in number of COVID-19 cases, this has also led to a spur in proposals on technology solutions for healthcare.
Specifically, the need for the development of simplistic, cost effective and fast testing, yet accurate methodologies for infection diagnosis has become  crucial for healthcare, policy making, and economic revival in several countries. The focus is also on point-of-care diagnostic tools, technology solutions which can be deployed rapidly, pre-screening tools, and cheaper alternatives to RT-PCR test, overcoming the limitations of chemical testing.

\begin{figure*}
\centering
\includegraphics[width=16cm,height=5cm]{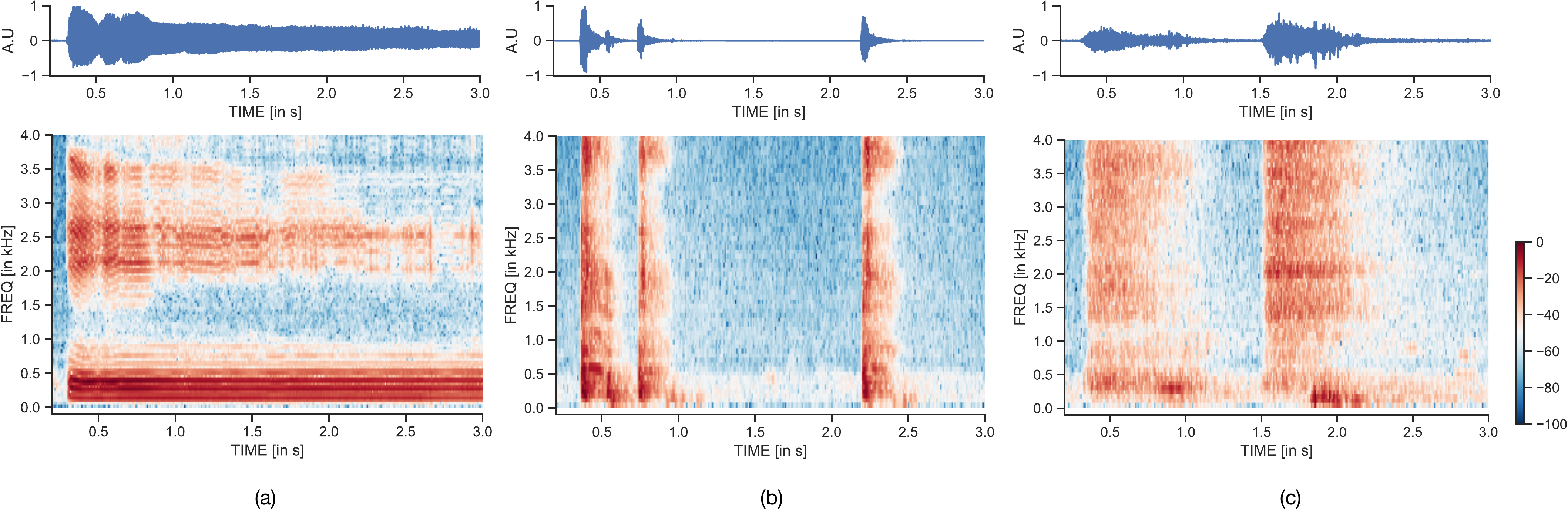}
\vspace*{-.25cm}
\caption{Spectrogram of recordings from a healthy  adult, (a) Sustained vowel \textit{/ey/} as in \textit{made}, (b) Heavy cough, and (c) Breathing: Inhalation followed by exhalation.}
\label{fig:sound_spectrum}
\end{figure*}

\noindent The research and the understanding of the novel virus and COVID-19 is a work in progress at various laboratories around the world. As of $16$ May, the WHO and the CDC have listed dry cough, difficulty in breathing, chest pain (or pressure), and loss of speech or movement as key symptoms of this viral infection, visible between $2-14$ days after exposure to the virus.
Also, a recent modeling study of symptoms data collected from a pool of $7178$ COVID-19 positive individuals validated the presence of these symptoms, and proposed a real-time prediction and tracking approach \cite{smell_metadata}.
Medical literature shows that speech breathing patterns are intricately tied to changes in anatomy and physiology of the respiratory system \cite{speech_breathing_book}.
Drawn by these observations, we identify an opportunity to make impact on point-of-care diagnosis using speech and acoustics research. Bringing together a large dataset of respiratory sounds, machine learning, and respiratory infection expertise from doctors can help in evaluating the potential in using respiratory sound samples for diagnosis of COVID-19.
The goal is not to replace the existing chemical testing methodologies but to supplement them with a cost effective, fast and simpler technique. This paper presents our efforts along this direction.

The project is named Coswara, a combination of the initials of coronavirus and swara (referring to sound in Sanskrit). The scientific rationale behind the conception of this project is presented in Section~\ref{sec:rationale}. Section~\ref{sec:project} presents an overview of the project. In Section~\ref{sec:dataset} a summary of the dataset collected and released openly as of $7$ August $2020$ is presented. We conclude with a discussion in Section~\ref{sec:conclusion}.

\section{Scientific Rationale}
\label{sec:rationale}
\subsection{Cough Sounds}
\vspace*{-.15cm}
Cough is a powerful reflex mechanism for the clearance of the central airways (trachea and the main stem bronchi) of inhaled and secreted material. Typically, it follows a well defined pattern, with an initial inspiration, glottal closure and development of high thoracic pressure, followed by an explosive expiratory flow as the glottis opens with continued expiratory effort \cite{thorpe2001acoustic}. The reflex is initiated by cough receptors within the central airways that are sensitive to both mechanical and chemical stimuli. Sound is generated during cough by air turbulence, vibration of the tissue, and movement of fluid through the airways \cite{polverino2012anatomy}. This turbulence generates sound with a broadband ``noisy'' character, whose frequency content depends on the velocity, density of the gas, and dimensions of the airways from source till the mouth. A cough sound is usually composed of three temporal phases: explosive, intermediate, and voicing phase. Fig.~\ref{fig:sound_spectrum}(b) shows a wide-band spectrogram of a sequence of three heavy cough sound signals. Each cough lasts close to $300$~ms, and the spectrum exhibits broad spectral spread over $500~$Hz, $1.5~$kHz, and $3.8~$kHz. Interestingly, over one hundred pathological conditions are associated with cough \cite{fontana2007cough}. The acoustic features of a cough sound depend on the air flow velocity as well as the dimensions of vocal tract and airways ~\cite{thorpe2001acoustic}.
This makes it possible to detect cough sounds \cite{cuong_cough_detection} in audio recordings.
As the physical structure of the respiratory system gets altered with respiratory infections, it is even possible to classify pathological condition based on a cough sound. Pertussis is a contagious respiratory disease which mainly affects young children and can be fatal if left untreated. Pramono et al.~\cite{pramono2016cough} presented an algorithm for automated diagnosis of pertussis using audio signals by analyzing cough
and whoop sounds. Several other recent studies have attempted to identify chronic obstructive pulmonary disease (COPD) (a disease caused primarily due to smoking) \cite{windmon2018tussiswatch}, and tuberculosis (an infectious disease usually caused by mycobacterium tuberculosis (MTB) bacteria that affects the lungs)~\cite{botha2018detection}. In addition, for respiratory disorders like asthma and pneumonia, algorithms based on cough sounds recorded using a smartphone provides a high level of accuracy \cite{porter2019prospective}.
For COVID-19 detection and diagnostics, the research initiatives from University of Cambridge \cite{covid19sounddetector}, Carnegie Mellon University \cite{covid19cmuproject}, Wadhwani AI institute \cite{coughagainstcovid20} and a project from EPFL \cite{coughvidepfl} have already been launched. Also, a recent work by Imran et al.~\cite{imran2020ai4covid} suggests a good accuracy for cough based detection of COVID-19 using a preliminary investigation with a small number of subjects.

\subsection{Breath sounds}
\vspace*{-.15cm}
Breathing difficulty is another common symptom for COVID-19. This is often exhibited as a shortness of breath. The early work by Anderson et al., \cite{anderson2001breath} reported that the spectrogram of breath sounds captured by a smartphone shows distinct patterns for asthmatic conditions compared to healthy individuals. 
For the diagnostics of COVID-19, the use of breathe sounds is also being attempted by a research group at New York University~\cite{breatheforscience20}. Fig.~\ref{fig:sound_spectrum}(c) shows the wide-band spectrograms of inhaling and exhaling cycles while breathing.

\subsection{Voice sounds}
\vspace*{-.15cm}
The studies show that lung diseases have distinct bio-markers in the speech breathing cycles \cite{lee1993speech}.  The study reported in \cite{chang2004perceived} showed that the phonation threshold pressure, defined as the minimal lung pressure required to initiate and sustain vocal fold oscillation, is correlated with vocal fatigue. The impact of laryngeal dysfunction in breathing patterns of read speech was analyzed in \cite{sapienza1997speech}. Fig.~\ref{fig:sound_spectrum}(a) depicts the spectrogram of a sustained phonation of /ey/ vowel (as in made). In contrast to cough and breath sound sample, there is a clear voicing seen as harmonics and localized concentration of energy at regions defined by formants. 

\section{Coswara Overview}
\label{sec:project}

\begin{figure*}[t!]
\centering
\includegraphics[width=14cm,height=6cm]{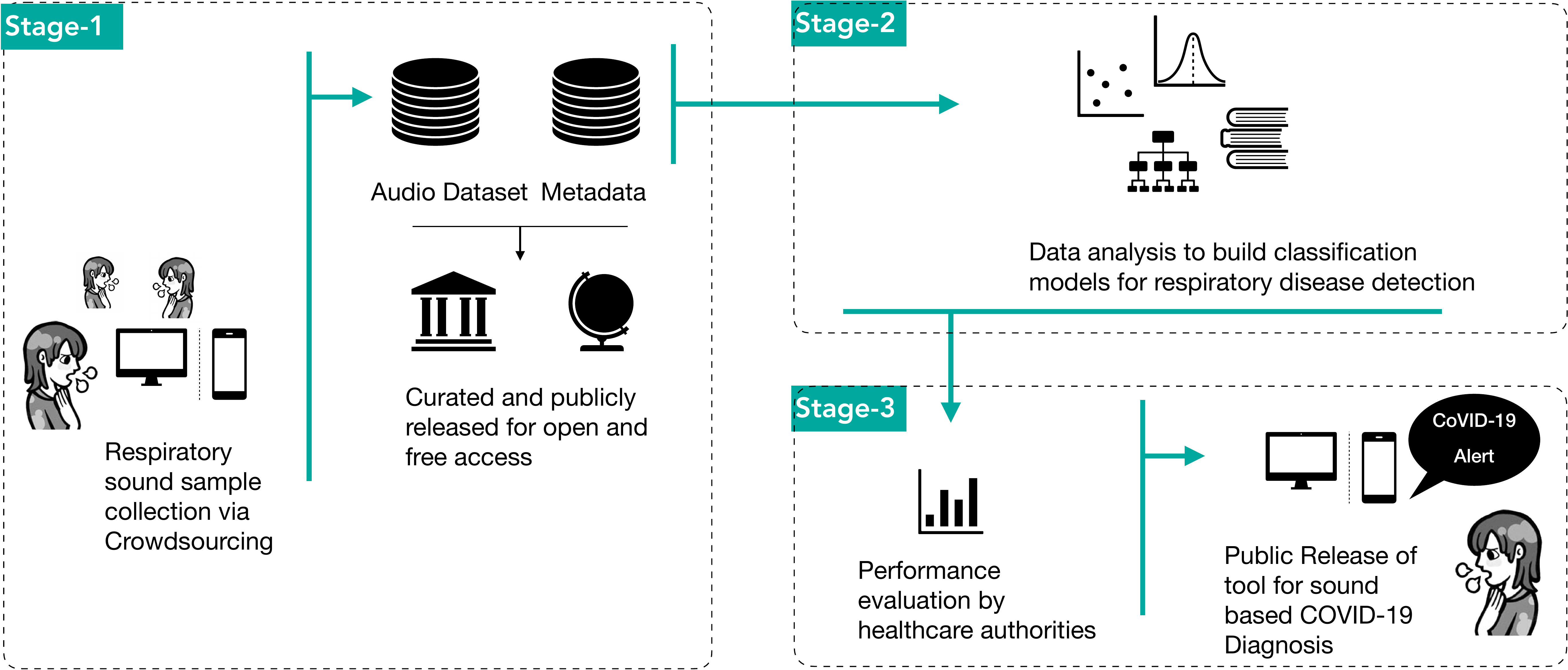}
\vspace{-0.1cm}
 \caption{Different stages in the proposed Coswara project - (1) Data collection, (2) Modeling, (3) Diagnostic tool development.}
\label{fig:stages}
\end{figure*}

Coswara~\cite{coswara20} is an attempt to provide a simple and cost effective tool for diagnosis of COVID-19 using breath, cough and speech sounds. As most of the major symptoms of the disease include respiratory problems, the proposed project aims to detect and quantify the bio-markers of the disease in the acoustics of the these sounds. The project has three stages, depicted in Fig. \ref{fig:stages}. Below we briefly describe each stage.

\subsection{Data collection}

\blfootnote{The web application for data collection is designed using HTML, and can be accessed at \url{https://coswara.iisc.ac.in/}.}The goal is to create a dataset of sound samples from healthy and unhealthy individuals, including those identified as COVID-19.
For sound data, we focus on nine different categories, namely, breathing (two kinds; shallow and deep), cough (two kinds; shallow and heavy), sustained vowel phonation (three kinds; /ey/~as in made, /i/~as in beet, /u:/ as in cool), and one to twenty digit counting (two kinds; normal and fast paced).
We also collect some metadata information, namely, age, gender, location (country, state/province), current health status (healthy / exposed / cured / infected) and the presence of co-morbidity (pre-existing medical conditions). No personally identifiable information is collected. The data is also anonymized during storage.
\subsection{Modeling}
\vspace*{-.15cm}
The collected data will be analysed using signal processing and machine learning techniques. The goal is to build mathematical models aiding identification of bio-markers from sound samples. This stage is a work-in-progress while we create the dataset. We have also initiated regular release of the curated dataset as open access via GitHub platform\footnote{The dataset and information about licensing can be accessed at \url{https://github.com/iiscleap/Coswara-Data}}.  

\subsection{Diagnosis tool}
\vspace*{-.15cm}
We aim to release the diagnosis tool as a web/mobile application. The application prompts the user to record voice samples, similar to the dataset collection stage, and provides a score indicating the probability of COVID-19 infection. The final deployment of tool is subject to validation with clinical findings, and authorization /approval from competent healthcare authorities. Given the highly simplistic and cost effective nature of the tool, we hypothesize that, even a partial success for the tool would enable a massive deployment as a first line diagnostic tool to pre-screen the infection.

\section{Dataset Description}
\label{sec:dataset}
\begin{figure*}[t!]
\centering
\includegraphics[width=16cm,height=7.5cm]{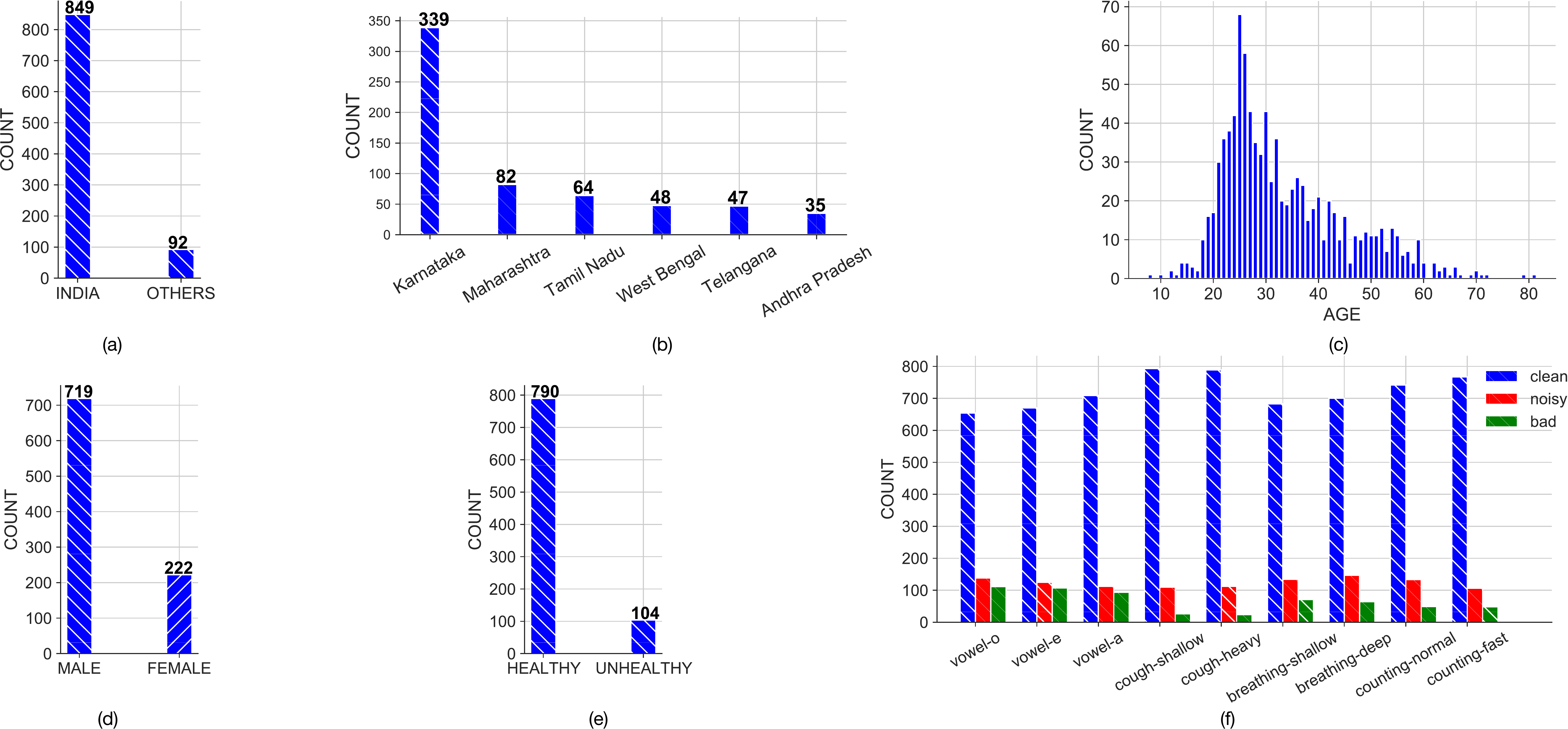}
\vspace*{-.25cm}
\caption{Metadata of the Coswara database: Participant count across (a) country, (b) Indian states, (c) age, (d) gender (only top $5$ are shown), (e) health status, and (f) sound category.}
\label{fig:metadata}
\end{figure*}
The project is currently on going. The collection, release, and analysis of the dataset are in progress. Below we provide a description of the most recent release of the dataset.

\subsection{Data collection methodology}
\vspace*{-.15cm}
The data collection strategy focused on reaching out to the human population across the globe. For this, we created a website application providing a simple and interactive user interface. A user could open the application in a web browser (laptop or mobile phone), provide metadata, and proceed to recording the sound samples using the device microphone. The average interaction time with the application is $5-7$~mins.
The user was prompted to use personal device and wipe off the device with a sanitizer before and after recording, and keep the device $10$~cm from the mouth during recording.

\subsection{Metadata description}
\vspace*{-.15cm}
The currently released dataset (as of 07 August 2020) has a participant count of $941$. Fig.~\ref{fig:metadata} shows the distribution of the participants across gender, age, country (grouped into India and outside), Indian states, and health status (grouped into healthy and unhealthy). Each participant provides $9$ audio files, one for each of the sound category.

\subsection{Annotation}
\begin{figure}[t!]
\centering
\includegraphics[width=8.0cm]{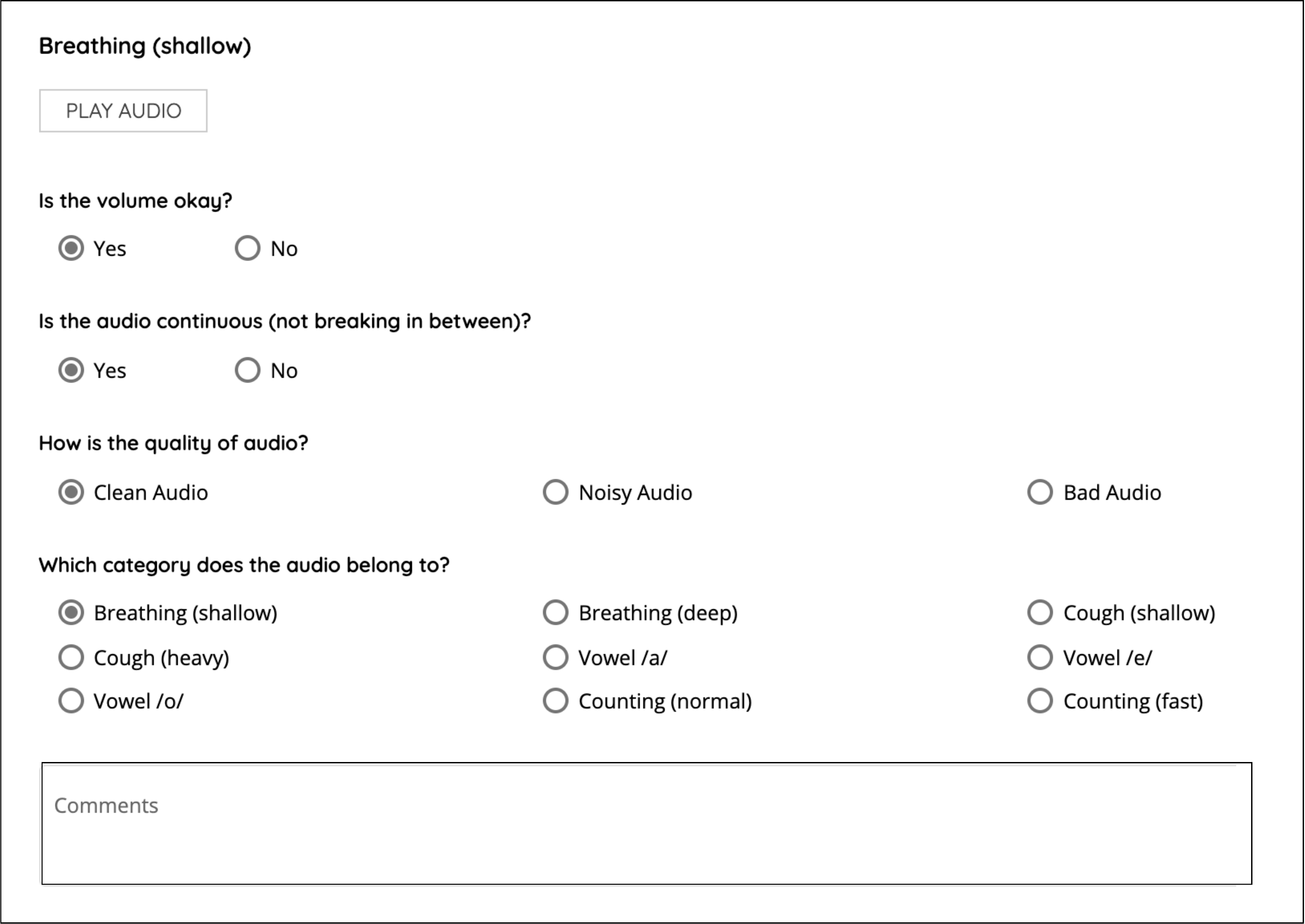}
\caption{A screenshot of the designed tool for annotation.}
\label{fig:annotator}
\vspace{-0.25cm}
\end{figure}

\vspace*{-.15cm}
The audio samples are recorded at a sampling frequency of $48~$kHz. All sound files were manually curated. A web interface was designed allowing the annotator (human) to listen to every sound file and answer some questions (a screenshot of the interface is shown in Fig.~\ref{fig:annotator}). These questions helped verify the category label, and the quality of the audio file. The annotator was also provide an option to provide any  additional comments for each audio file. Fig.~\ref{fig:metadata}(f) depicts the quality count of the recordings, pooling all the nine sound categories. Majority of the audio files were rated as clean. While we had $13$ annotators, each file was listened by only one annotator. In total, the dataset has $6507$ clean, $1117$ noisy, and remaining highly degraded audio files corresponding to respiratory sound samples from $941$ participants.

\begin{figure}[t!]
\centering
\includegraphics[width=8.5cm,height=7cm]{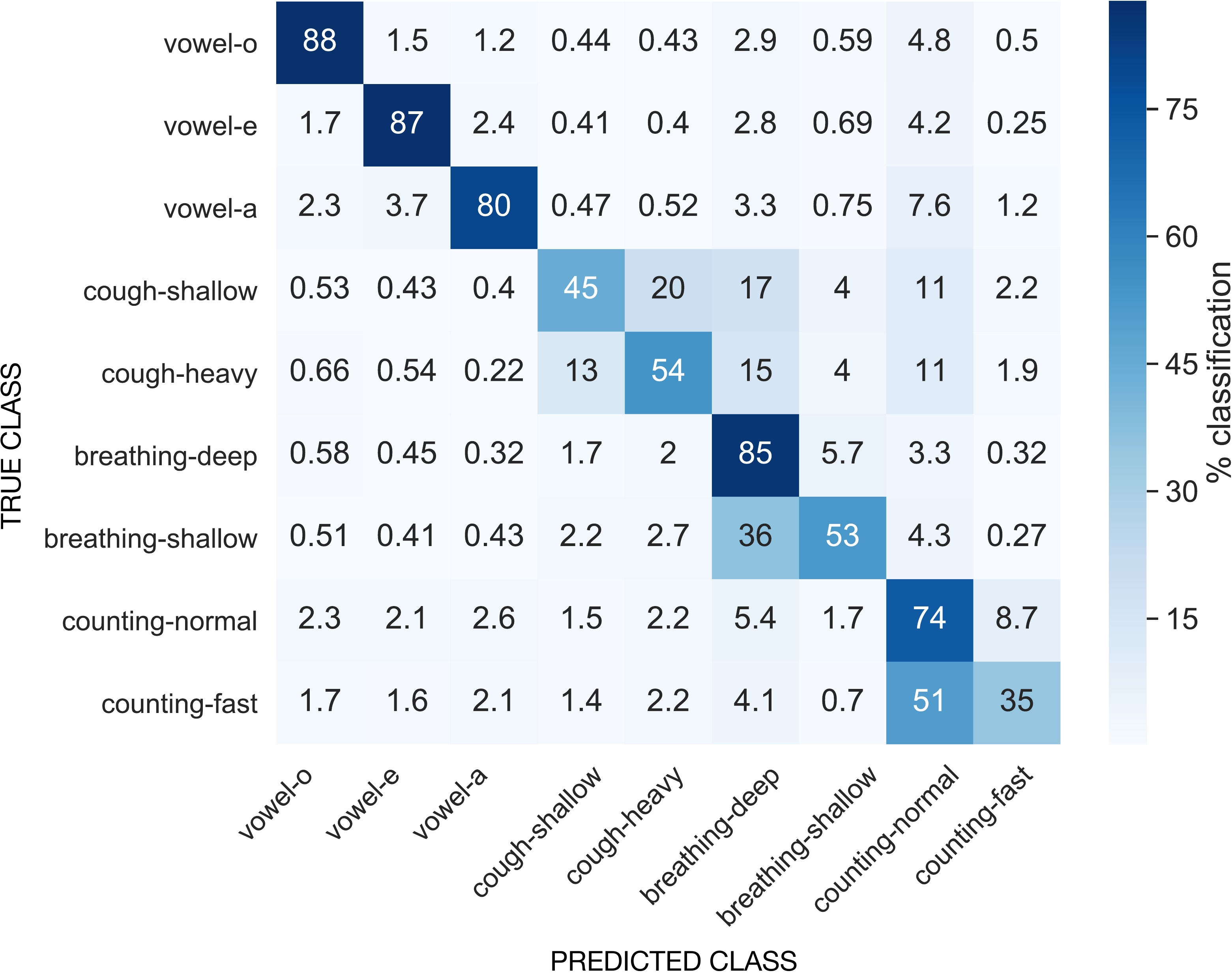}
\vspace{-0.5cm}
\caption{Confusion matrix (on test data) obtained on classifying 9 categories of sounds using a random forest classifier.}
\label{fig:9class}
\vspace*{-0.5cm}
\end{figure}

\subsection{Acoustic Properties}
The $9$ sound categories (or classes) are chosen such that physical state of the respiratory system is well captured just by using the sound samples. We tested the complementarity across these sound categories by building a multi-class classifier trained and tested on acoustic features extracted from the different sound samples. The goal was to build a $9$-class (corresponding to the $9$ sound categories) classifier and evaluate the confusion matrix.

The clean audio recordings were pooled and grouped by $9$ sound categories. A set of different short-time ($500$~msec, with hop of $100~$msec) temporal and spectral acoustic features were extracted from the audio files. These included spectral contrast ($7$-D), MFCCs ($13$-D), spectral roll-off ($1$-D), spectral centroid ($1$-D), mean square energy ($1$-D), polynomial fit to the spectrum ($2$-D), zero-crossing rate ($1$-D), spectral bandwidth ($1$-D), and spectral flatness ($1-$D). After concatenation each $500$~msec segment of audio was represented by a $28$-D feature. A random forest classifier (see \cite{scikit-learn}, built with default parameters and $30$ trees) was trained on a $70-30$\% train-test split for classifying every $500$~msec segment into one of the $9$ sound categories.

The accuracy on test data was $66.74\%$. Fig.~\ref{fig:9class} shows resulting confusion matrix for the test data. The test data had an equal share from every class. It is interesting to note that the vowels are less confused, and the digit counting samples are more confused between themselves. This is also the case for cough samples and breathing samples.

\section{Conclusion}
\label{sec:conclusion}
We have described our efforts towards a sound based diagnostic tool for COVID-19. The tool, named Coswara, built upon prior studies that have shown good accuracy for detecting other respiratory disorders like asthma, pertussis, tuberculosis and pneumonia. We highlight the rationale for choosing different stimuli in the Coswara database. The progress in data collection in terms of meta data statistics are described. We also highlight the complimentary nature of the stimuli chosen. The next phase in the diagnostic tool development will attempt the use of machine learning algorithms to classify different health conditions and try to identify sound based bio-markers for Covid-19. 

\section{Acknowledgements}
We thank Anand Mohan for his enormous help in web design and data collection efforts, which allowed the recording of data from general public. We also thank Chandrika Thunga, Sonali Singh, Sakya Basak, Venkat Krishnamohan, Jaswanth Reddy, Ananya Muguli, Anirudh Sreeram and Anagha Rajeev for their contribution in annotating the dataset.

\bibliographystyle{IEEEtran}

\bibliography{mybib}

\begin{thebibliography}{10}
\providecommand{\url}[1]{#1}
\csname url@samestyle\endcsname
\providecommand{\newblock}{\relax}
\providecommand{\bibinfo}[2]{#2}
\providecommand{\BIBentrySTDinterwordspacing}{\spaceskip=0pt\relax}
\providecommand{\BIBentryALTinterwordstretchfactor}{4}
\providecommand{\BIBentryALTinterwordspacing}{\spaceskip=\fontdimen2\font plus
\BIBentryALTinterwordstretchfactor\fontdimen3\font minus
  \fontdimen4\font\relax}
\providecommand{\BIBforeignlanguage}[2]{{%
\expandafter\ifx\csname l@#1\endcsname\relax
\typeout{** WARNING: IEEEtran.bst: No hyphenation pattern has been}%
\typeout{** loaded for the language `#1'. Using the pattern for}%
\typeout{** the default language instead.}%
\else
\language=\csname l@#1\endcsname
\fi
#2}}
\providecommand{\BIBdecl}{\relax}
\BIBdecl

\bibitem{world2020naming}
``Naming the coronavirus disease ({COVID-19}) and the virus that causes it,''
  \emph{World Health Organization. https://www.who.
  int/emergencies/diseases/novel-coronavirus-2019/technical-guidance/naming-the-coronavirus-disease-(covid-2019)-and-the-virus-that-causes-it},
  2020.

\bibitem{smell_metadata}
\BIBentryALTinterwordspacing
C.~Menni, A.~M. Valdes, M.~B. Freidin, C.~H. Sudre, L.~H. Nguyen, D.~A. Drew,
  S.~Ganesh, T.~Varsavsky, M.~J. Cardoso, J.~S. El-Sayed~Moustafa, A.~Visconti,
  P.~Hysi, R.~C.~E. Bowyer, M.~Mangino, M.~Falchi, J.~Wolf, S.~Ourselin, A.~T.
  Chan, C.~J. Steves, and T.~D. Spector, ``Real-time tracking of self-reported
  symptoms to predict potential {COVID-19},'' \emph{Nature Medicine}, 2020.
  [Online]. Available: \url{https://doi.org/10.1038/s41591-020-0916-2}
\BIBentrySTDinterwordspacing

\bibitem{speech_breathing_book}
J.~E. Huber and E.~T. Stathopoulos, \emph{Speech Breathing Across the Life Span
  and in Disease}.\hskip 1em plus 0.5em minus 0.4em\relax John Wiley \& Sons,
  Ltd, 2015, ch.~2, pp. 11--33.

\bibitem{thorpe2001acoustic}
W.~Thorpe, M.~Kurver, G.~King, and C.~Salome, ``Acoustic analysis of cough,''
  in \emph{The Seventh Australian and New Zealand Intelligent Information
  Systems Conference}.\hskip 1em plus 0.5em minus 0.4em\relax IEEE, 2001, pp.
  391--394.

\bibitem{polverino2012anatomy}
M.~Polverino, F.~Polverino, M.~Fasolino, F.~And{\`o}, A.~Alfieri, and
  F.~De~Blasio, ``Anatomy and neuro-pathophysiology of the cough reflex arc,''
  \emph{Multidisciplinary respiratory medicine}, vol.~7, no.~1, p.~5, 2012.

\bibitem{fontana2007cough}
G.~A. Fontana and J.~Widdicombe, ``What is cough and what should be measured?''
  \emph{Pulmonary pharmacology \& therapeutics}, vol.~20, no.~4, pp. 307--312,
  2007.

\bibitem{cuong_cough_detection}
C.~Pham, ``Mobicough: Real-time cough detection and monitoring using low-cost
  mobile devices,'' in \emph{Intelligent Information and Database Systems},
  N.~T. Nguyen, B.~Trawi{\'{n}}ski, H.~Fujita, and T.-P. Hong, Eds.\hskip 1em
  plus 0.5em minus 0.4em\relax Berlin, Heidelberg: Springer Berlin Heidelberg,
  2016, pp. 300--309.

\bibitem{pramono2016cough}
R.~X.~A. Pramono, S.~A. Imtiaz, and E.~Rodriguez-Villegas, ``A cough-based
  algorithm for automatic diagnosis of pertussis,'' \emph{PloS one}, vol.~11,
  no.~9, 2016.

\bibitem{windmon2018tussiswatch}
A.~Windmon, M.~Minakshi, P.~Bharti, S.~Chellappan, M.~Johansson, B.~A. Jenkins,
  and P.~R. Athilingam, ``Tussiswatch: A smart-phone system to identify cough
  episodes as early symptoms of chronic obstructive pulmonary disease and
  congestive heart failure,'' \emph{IEEE journal of biomedical and health
  informatics}, vol.~23, no.~4, pp. 1566--1573, 2018.

\bibitem{botha2018detection}
G.~Botha, G.~Theron, R.~Warren, M.~Klopper, K.~Dheda, P.~Van~Helden, and
  T.~Niesler, ``Detection of tuberculosis by automatic cough sound analysis,''
  \emph{Physiological measurement}, vol.~39, no.~4, p. 045005, 2018.

\bibitem{porter2019prospective}
P.~Porter, U.~Abeyratne, V.~Swarnkar, J.~Tan, T.-w. Ng, J.~M. Brisbane,
  D.~Speldewinde, J.~Choveaux, R.~Sharan, K.~Kosasih \emph{et~al.}, ``A
  prospective multicentre study testing the diagnostic accuracy of an automated
  cough sound centred analytic system for the identification of common
  respiratory disorders in children,'' \emph{Respiratory research}, vol.~20,
  no.~1, p.~81, 2019.

\bibitem{covid19sounddetector}
``Cambridge university, {UK} - {COVID-19} sounds app,'' \emph{Weblink
  https://covid-19-sounds.org/en/}, last accessed May 8, 2020.

\bibitem{covid19cmuproject}
``{CMU} sounds for covid project,'' \emph{Weblink
  https://node.dev.cvd.lti.cmu.edu/}, last accessed May 8, 2020.

\bibitem{coughagainstcovid20}
``Cough against covid - wadhwani {AI} institute,'' \emph{Weblink
  https://coughagainstcovid.org/}, 2020.

\bibitem{coughvidepfl}
``Cough for {COVID-19} detection,'' \emph{Weblink https://coughvid.epfl.ch/},
  last accessed May 8, 2020.

\bibitem{imran2020ai4covid}
A.~Imran, I.~Posokhova, H.~N. Qureshi, U.~Masood, S.~Riaz, K.~Ali, C.~N. John,
  and M.~Nabeel, ``{AI4COVID-19: AI} enabled preliminary diagnosis for
  {{COVID-19}} from cough samples via an app,'' \emph{arXiv preprint
  arXiv:2004.01275}, 2020.

\bibitem{anderson2001breath}
K.~Anderson, Y.~Qiu, A.~R. Whittaker, and M.~Lucas, ``Breath sounds, asthma,
  and the mobile phone,'' \emph{The Lancet}, vol. 358, no. 9290, pp.
  1343--1344, 2001.

\bibitem{breatheforscience20}
``Breathing sounds for {COVID-19},'' \emph{Weblink
  https://breatheforscience.com/}, last accessed May 8, 2020.

\bibitem{lee1993speech}
L.~Lee, R.~G. Loudon, B.~H. Jacobson, and R.~Stuebing, ``Speech breathing in
  patients with lung disease,'' \emph{American Review of Respiratory Disease},
  vol. 147, pp. 1199--1199, 1993.

\bibitem{chang2004perceived}
A.~Chang and M.~P. Karnell, ``Perceived phonatory effort and phonation
  threshold pressure across a prolonged voice loading task: a study of vocal
  fatigue,'' \emph{Journal of Voice}, vol.~18, no.~4, pp. 454--466, 2004.

\bibitem{sapienza1997speech}
C.~M. Sapienza, E.~T. Stathopoulos, and W.~Brown~Jr, ``Speech breathing during
  reading in women with vocal nodules,'' \emph{Journal of Voice}, vol.~11,
  no.~2, pp. 195--201, 1997.

\bibitem{coswara20}
``Indian institute of science - {C}oswara: A sound based diagnostic tool for
  covid19,'' \emph{Weblink https://coswara.iisc.ac.in/}, last accessed May 8,
  2020.

\end{thebibliography}
\end{document}